\documentclass[aps,showpacs,preprintnumbers,amsmath,amssymb]{revtex4}

\usepackage{float}
\usepackage{graphics,epsfig}
\usepackage{graphicx}
\usepackage{dcolumn}
\usepackage{bm}
\newcommand{\no}{\nonumber}

\begin{document}
\baselineskip=0.8 cm
\title{{\bf Scalar emission in a rotating G\"{o}del black hole}}
\author{Songbai Chen}
\email{csb3752@163.com} \affiliation{ Institute of Physics and
Department of Physics, Hunan Normal University,  Changsha, Hunan
410081, P. R. China \\ Key Laboratory of Low Dimensional Quantum
Structures \\ and Quantum Control of Ministry of Education, Hunan
Normal University, Changsha, Hunan 410081, P. R. China
\\ Department of Physics, Fudan University,
Shanghai 200433, P. R. China}

\author{Bin Wang}
\email{wangb@fudan.edu.cn} \affiliation{Department of Physics,
Fudan University, Shanghai 200433, P. R. China}

\author{Jiliang Jing}
\email{jljing@hunnu.edu.cn}
 \affiliation{Institute of Physics and
Department of Physics, Hunan Normal University,  Changsha, Hunan
410081, P. R. China \\ Key Laboratory of Low Dimensional Quantum
Structures \\ and Quantum Control of Ministry of Education, Hunan
Normal University, Changsha, Hunan 410081, P. R. China}
\begin{abstract}
\baselineskip=0.6 cm
\begin{center}
{\bf Abstract}
\end{center}

We study the absorption probability and Hawking radiation of the
scalar field in the rotating G\"{o}del black hole in minimal
five-dimensional gauged supergravity. We find that G\"{o}del
parameter $j$ imprints in the greybody factor and Hawking
radiation. It plays a different role from the angular momentum of
the black hole in the Hawking radiation and super-radiance. These
information can help us know more about rotating G\"{o}del black
holes in minimal five-dimensional gauged supergravity.

\end{abstract}

 \maketitle
\newpage
\section{Introduction}

The standard Friedmann-Robertson-Walker (FRW) model is a popular
phenomenological description of the rather idealized isotropic and
homogeneous universe filled with perfect fluid. However this model
is too ideal to be used to describe the universe with global
rotation. Considering that all compact objects in the universe are
rotating, it would be nature to think that the rotation is a
universal phenomenon which may also apply to the global universe. An
exact solution for the rotating universe was found by G\"{o}del
\cite{godel}, which satisfies Einstein's field equations containing
a cosmological constant and homogeneous pressureless matter.
G\"{o}del universe possesses several special features including, in
particular, the presence of closed timelike curve (i.e. the time
machine) \cite{Biswas,Brecher} through every point.

A great deal of effort
\cite{jpb,car,duh,eks,ttt,bh1,DKL,GGc,MGW,RlM,DRB1} has been spent
in recent years in studying G\"{o}del-type solutions in the context
of five-dimensional minimal supergravity. These solutions are
related by T duality to pp-waves \cite{eks,ttt}, which make the
G\"{o}del-type universes important since they might provide the
possibility of quantizing strings in these backgrounds and building
relations to the corresponding limits of super-Yang Mills theories.

The black hole solution must embed in the G\"{o}del universe. The
solutions of the neutral non-rotating and rotating black holes
immersed in the rotating G\"{o}del universe were found in the
five-dimensional minimal supergravity \cite{bh1}, which are called
Schwarzschild and Kerr G\"{o}del black holes. These black hole
solutions also satisfy the usual black hole thermodynamics
\cite{DKL,GGc,MGW,RlM,DRB1}. In the study of the wave dynamical
properties of Schwarzschild G\"{o}del black hole\cite{qu1}, it was
found that the Schwarzschild G\"{o}del black hole is stable at least
for the small G\"{o}del parameter regime. Recently, there have been
more solutions of black holes found in G\"{o}del universe and
various properties have been investigated accordingly
\cite{ss1,ss2,ss3,ss4}. In this paper we are going to study the
Hawking radiation in the rotating black hole embedded in the
G\"{o}del universe. We will calculate the greybody factor of scalar
particles propagating in this Kerr G\"{o}del black hole and show
rich physics brought by the rotation parameter $a$ of the black hole
and the G\"{o}del parameter $j$ by using the matching technique,
which has been extensively used in evaluating the absorption
probabilities and Hawking radiations of various black holes
\cite{Kanti,Kan1,Kan2,Kan3,Kan4,Kan5,Kan6,Haw3,Haw4,Haw5}.

The paper is organized as follows: in the following section we
will introduce the Kerr G\"{o}del metric and derive the master
equation of scalar field in the limit of small G\"{o}del parameter
$j$. In Sec.III, we will obtain the approximate solution in the
low energy and low angular momentum limit. In section IV, we will
calculate the absorption probability and the luminosity of Hawking
radiation for the scalar field. Finally in the last section we
will include our conclusions.

\section{Master equation in the Kerr G\"{o}del Black Holes}

The equations of the bosonic field in the minimal supergravity
theory in five dimensions are written in the form
\begin{eqnarray}
&&R_{\mu\nu}=2F_{\mu\alpha}F_{\nu}^{\;\alpha}-\frac{1}{6}g_{\mu\nu}F^2,\nonumber\\
&&D_{\mu}F^{\mu\nu}=\frac{1}{2\sqrt{3}}\;\tilde{\epsilon}^{\alpha\beta\gamma\mu\nu}F_{\alpha\beta}F_{\gamma\mu},
\label{go1}
\end{eqnarray}
where
$\tilde{\epsilon}_{\alpha\beta\gamma\mu\nu}=\sqrt{-g}\epsilon_{\alpha\beta\gamma\mu\nu}$
and $F^2=F_{\alpha\beta}F^{\alpha\beta}$. The solutions of equations
(\ref{go1}) consist of the metric and a $U(1)$ gauge field which has
an additional Chern-Simons interaction.

The G\"{o}del universe is a solution of the equation (\ref{go1}),
whose metric and $U(1)$ gauge field are given by \cite{bh1}
\begin{eqnarray}
ds^2&=&-(dt+jr^2\sigma_3)^2+dr^2+\frac{r^2}{4}d\Omega^2_3,\label{m1}
\;\;\;\;\;\;\tilde{A}=\frac{\sqrt{3}}{2}jr^2\sigma_3,
\end{eqnarray}
with
\begin{eqnarray}
d\Omega^2_3=d \theta^2+\sin^2{\theta}d \phi^2+\sigma^2_3,\;\;\;\;
\sigma_3=d \psi+\cos{\theta}d \phi,
\end{eqnarray}
where $\theta$, $\phi$ and $\psi$ are Euler angles. $j$ is the
G\"{o}del parameter and is responsible for the rotation of the
G\"{o}del universe. The angular velocity of the universe is
$\Omega_u=\frac{4j}{1-4j^2r^2}$. When $r>1/(2j)$, the closed
timelike curves appear. The matter content of G\"{o}del universe
(\ref{m1}) consists of pressureless dust and the energy-momentum
tensor for the field $F_{\mu\nu}$ has vanishing pressure and
constant energy density proportional to $j^2$. In addition, just
like the original four dimensional G\"{o}del universe \cite{godel},
the solution (\ref{m1}) is homogeneous. When $j=0$, the metric
(\ref{m1}) reduces to the Minkowski spacetime.

The solution of the equation (\ref{go1}) which describes a five
dimensional rotating black hole (with the two equal rotation
parameters) embedded in the G\"{o}del universe can be written as
\cite{bh1}
\begin{eqnarray}
ds^2&=&-u(r)dt^2-2g(r)\sigma_3dt+h(r)\sigma^2_3+f(r)^{-1}dr^2+\frac{r^2}{4}d\Omega^2_3,\label{metric0}
\end{eqnarray}
with
\begin{eqnarray}
u(r)&=&1-\frac{2M}{r^2},\;\;\;\;\;\;\;\;\;h(r)=-j^2r^2(r^2+2m)+\frac{Ma^2}{2r^2} ,\nonumber\\
g(r)&=&jr^2+\frac{Ma}{r^2},\;\;\;\;\;f(r)=1-\frac{2M}{r^2}+\frac{8jM(a+2jM)}{r^2}+\frac{2Ma^2}{r^4}.
\end{eqnarray}
Here $M$ and $a$ are related to the mass and angular momentum of
the black hole, respectively. When $M=a=0$, the metric reduces to
that of the five-dimensional G\"{o}del universe (\ref{m1}). When
$j=0$ it reduces to the five-dimensional Kerr black hole with two
equal rotation parameters. When $a=0$ the solution becomes the
Schwarzschild-G\"{o}del black hole.

The Kerr G\"{o}del Black Hole (\ref{metric0}) has an outer horizon
at $r_+$ and an inner horizon at $r_-$,
\begin{eqnarray}
r^2_{\pm}=M(1-4aj-8j^2M)\pm\sqrt{M^2(1-4aj-8j^2M)^2-2Ma^2}.
\end{eqnarray}
which are determined by $f(r)=0$. The metric is well behaved at
the horizon but the gauge field becomes singular there. The
function $u(r)$ is equal to zero when $r=\sqrt{2M}$, corresponding
to an ergosphere \cite{RlM}. The Hawking temperature $T_H$ of the
black hole and the angular velocity $\Omega_H$ at horizon are
described as
\begin{eqnarray}
T_H=\frac{r^2_+-r^2_-}{r^2_+\sqrt{4h(r_+)+r^2_+}},\label{T1}
\end{eqnarray}
and
\begin{eqnarray}
\Omega_H=\frac{4g(r_+)}{4h(r_+)+r^2_+},\label{Omega1}
\end{eqnarray}
respectively.

In the following we will investigate Hawking radiation of a scalar
field in the small $j$ case since the small rotation of the
G\"{o}del cosmological background seems the most reasonable in
phenomenology \cite{qu1}. In the limit of small $j$, the metric
coefficients (\ref{metric0}) can be rewritten as
\begin{eqnarray}
u(r)&=&1-\frac{2M}{r^2},\;\;\;\;\;\;\;\;\;h(r)=\frac{Ma^2}{2r^2} ,\nonumber\\
g(r)&=&jr^2+\frac{Ma}{r^2},\;\;\;\;\;f(r)=1-\frac{2M}{r^2}+\frac{8jMa}{r^2}+\frac{2Ma^2}{r^4}.
\end{eqnarray}
The non-zero components of the inverse metric $g^{\mu\nu}$ and the
determinant $g$ can be expressed as
\begin{eqnarray}
g^{00}&=&-\frac{r^4+2Ma^2}{\Delta},\;\;\;\;\;\;\;\;\;\;\;\;\;\;\;\;\;\;\;\;\;\;\;\;g^{11}=\frac{\Delta}{r^4},
\;\;\;\;\;\;\;\;\;\;\;\;\;\;\;\;\;\;\;\;\;\;\;\;\;\;\;\;g^{22}=\frac{4}{r^2},\nonumber\\
g^{04}&=&-\frac{4(jr^4+Ma)}{\Delta},\;\;\;\;\;\;\;\;\;\;\;\;\;\;\;\;\;\;\;\;
g^{34}=-\frac{4\cos{\theta}}{r^2\sin^2{\theta}},\;\;\;\;\;\;\;\;\;\;\;\;\;\;\;\;\;
g^{33}=\frac{4}{r^2\sin^2{\theta}},\nonumber\\
g^{44}&=&\frac{4\cos^2{\theta}}{r^2\sin^2{\theta}}+\frac{4(r^2-2M)}{\Delta},
\;\;\;\;\;\;\;\;\;\;g=-\frac{r^6}{64}\sin^2{\theta},
\end{eqnarray}
with
\begin{eqnarray}
\Delta=r^4-2M(1-4aj)r^2+2Ma^2.
\end{eqnarray}

The wave equation for the massless scalar field
$\Phi(t,r,\theta,\phi,\psi)$ in the Kerr G\"{o}del black hole
spacetime obeys
\begin{eqnarray}
\frac{1}{\sqrt{-g}}\partial_{\mu}\bigg(\sqrt{-g}g^{\mu\nu}\partial_{\nu}
\Phi(t,r,\theta,\phi,\psi)\bigg)=0.\label{WE}
\end{eqnarray}
Taking the ansatz $\Phi(t,r,\theta,\phi,\psi)=e^{-i\omega t}R(r)e^{i
m\phi+i\lambda \psi}S(\theta)$, where $S(\theta)$ is the so-called
spheroidal harmonics, we can obtain the equation
\begin{eqnarray}
\frac{1}{\sin{\theta}}\frac{d}{d\theta}\bigg[\sin{\theta} \frac{d
S(\theta)}{d \theta}\bigg]
-\bigg[\frac{(m-\lambda\cos{\theta})^2}{\sin^2{\theta}}-E_{lm\lambda}\bigg]S(\theta)=0,\label{angd}
\end{eqnarray}
for the angular part. Obviously, this angular equation is
independent of the rotating parameters $a$ and $j$, and is exactly
identical to that in the static five-dimensional black hole
spacetime. This is not surprising and in \cite{sq6} the same
angular equation as that in the Schwarzschild spacetime was also
obtained in the five-dimensional Kerr black hole with two equal
rotational parameters. The eigenvalue of the angular equation
(\ref{angd}) is $E_{lm\lambda}=l(l+1)-\lambda^2$. The radial part
reads
\begin{eqnarray}
\frac{1}{r^3}\frac{d}{dr}\bigg[\frac{\Delta}{r}\frac{d
R(r)}{dr}\bigg]
+\bigg[\frac{K^2}{\Delta}-\frac{4E_{lm\lambda}+\Lambda}{r^2}\bigg]R(r)=0,\label{radial}
\end{eqnarray}
with
\begin{eqnarray}
K=\sqrt{r^4+2Ma^2}\bigg[\omega-\frac{4\lambda
(jr^4+Ma)}{r^4+2Ma^2}\bigg],\;\;\;\;\;\;\;\;\;\;\;\;\;\;\;
\Lambda=\frac{4\lambda^2r^4}{r^4+2Ma^2}.
\end{eqnarray}
The solution of the radial function $R(r)$ will help us to obtain
the absorption probability $|\mathcal{A}_{lm\lambda}|^2$ and the
luminosity of Hawking radiation for a massless scalar particle
propagating in the black hole spacetime.

\section{Greybody Factor in the Low-Energy Regime}

Now we try to obtain an analytic solution of the radial equation
(\ref{radial}) by using the well-known approximation technique
\cite{Haw3}: we first solve the equation in the near horizon regime
($r\simeq r_+$) and then in the far field limit $(r\gg r_+)$,
finally we smoothly match these two solutions in an intermediate
region. In this way we can get an analytic expression in the low
energy and low angular momentum approximation for the radial part of
the field valid throughout the whole spacetime.

Let us first focus on the near-horizon regime. In order to express
equation (\ref{radial}) into the form of a known differential
equation,  we perform the following transformation of the radial
variable \cite{Haw3}
\begin{eqnarray}
r\rightarrow f=\frac{\Delta}{r^4}\Rightarrow \frac{d
f}{dr}=(1-f)\frac{A}{r},
\end{eqnarray}
where
\begin{eqnarray}
A=2\bigg[1-\frac{a^2}{(1-4aj)r^2-a^2}\bigg].\label{ax}
\end{eqnarray}
The equation (\ref{radial}) near the horizon $(r\sim r_+)$ can be
expressed as
\begin{eqnarray}
f(1-f)\frac{d^2R(f)}{d f^2}+(1-D_*f)\frac{d R(f)}{d f}
+\bigg\{\frac{K^2_*}{A(\rho_+)^2(1-f)f}
-\frac{4E_{lm\lambda}+\Lambda(r_+)}{A(r_+)^2(1-f)}\bigg\}R(f)=0,\label{r1}
\end{eqnarray}
where
\begin{eqnarray}
K_*&=&\sqrt{1+\frac{2Ma^2}{r^2_+}}r_+\bigg[\omega-\frac{4\lambda(jr^4_++Ma)}{r^4_++2Ma^2}\bigg],\nonumber\\
D_*&=&1-2a^2\frac{[(1-4aj)r^2_+-a^2]}{[(1-4aj)r^2_+-2a^2]^2}.\label{kx}
\end{eqnarray}
Employing the transformation $R(f)=f^{\alpha}(1-f)^{\beta}F(f)$,
we can write the equation (\ref{r1}) into the form of the
hypergeometric equation
\begin{eqnarray}
f(1-f)\frac{d^2F(f)}{d f^2}+[c-(1+a_1+b)f]\frac{d F(f)}{d f}-a_1b
F(f)=0,\label{r2}
\end{eqnarray}
with
\begin{eqnarray}
a_1=\alpha+\beta+D_*-1,\;\;\;\;\;\;\;\;\;\;
b=\alpha+\beta,\;\;\;\;\;\;\;\;\;\;\;\;\; c=1+2\alpha.
\end{eqnarray}
Due to the constraint from coefficient of $F(f)$, the power
coefficients $\alpha$ and $\beta$ must satisfy the second-order
algebraic equations
\begin{eqnarray}
\alpha^2+\frac{K^2_*}{A(r_+)^2}=0,
\end{eqnarray}
and
\begin{eqnarray}
\beta^2+\beta(D_*-2)+\frac{K^2_*}{A(r_+)^2}-
\frac{4E_{lm\lambda}+\Lambda(r_+)}{A(r_+)^2}=0,
\end{eqnarray}
respectively. Solving these two equations, we obtain the solutions
for $\alpha$ and $\beta$
\begin{eqnarray}
&&\alpha_{\pm}=\pm \frac{iK_*}{A(r_+)},\\
&&\beta_{\pm}=\frac{1}{2}\bigg[(2-D_*)\pm\sqrt{(D_*-2)^2-\frac{4K^2_*}{A(r_+)^2}
+\frac{4[4E_{lm\lambda}+\Lambda(r_+)]}{A(r_+)^2}}
\;\bigg].\label{bet}
\end{eqnarray}
The general solution of the master equation (\ref{radial}) near the
horizon can be expressed as
\begin{eqnarray}
R_{NH}(f)=A_-f^{\alpha}(1-f)^{\beta}F(a_1,b,c;
f)+A_+f^{-\alpha}(1-f)^{\beta}F(a_1-c+1,b-c+1,2-c; f),\label{s0}
\end{eqnarray}
where $A_{\pm}$ are arbitrary constants. Considering the boundary
condition that no outgoing mode exists near the horizon, we choose
$\alpha=\alpha_-$ and $A_+=0$, which brings the near horizon
solution to the final form
\begin{eqnarray}
R_{NH}(f)=A_-f^{\alpha}(1-f)^{\beta}F(a_1, b, c; f).
\end{eqnarray}
Moreover, the above boundary condition also demands that near the
horizon the hypergeometric function $F(a_1, b, c; f)$ must be
convergent, i.e. $Re(c-a_1- b)> 0$, which implies that we must
choose $\beta=\beta_-$.

In order to construct a full analytic solution valid for the whole
radial regime, we need to smoothly match the near horizon and far
field solutions in the intermediate zone. For the benefit of the
following discussion, we change the expression of the
hypergeometric function of the near horizon solution from $f$ to
$1-f$ by using the relation
\begin{eqnarray}
R_{NH}(f)&=&A_-f^{\alpha}(1-f)^{\beta}\bigg[\frac{\Gamma(c)\Gamma(c-a_1-b)}{\Gamma(c-a_1)\Gamma(c-b)}
F(a_1, b, a_1+b-c+1; 1-f)\nonumber\\
&+&(1-f)^{c-a_1-b}\frac{\Gamma(c)\Gamma(a_1+b-c)}{\Gamma(a_1)\Gamma(b)}
F(c-a_1, c-b, c-a_1-b+1; 1-f)\bigg],\label{r2}
\end{eqnarray}
and stretch it towards the intermediate regime. In the limit $r\gg
r_+$, the function $(1-f)$ can be approximated as
\begin{eqnarray}
1-f\simeq \frac{2M(1-4aj)}{r^2},
\end{eqnarray}
and the near horizon solution (\ref{r2}) can be simplified further
to
\begin{eqnarray}
R_{NH}(r)\simeq A_1r^{-2\beta}+A_2r^{2(\beta+D_*-2)}\label{rn2},
\end{eqnarray}
and
\begin{eqnarray}
A_1=A_-[2M(1-4aj)]^{\beta}
\frac{\Gamma(c)\Gamma(c-a_1-b)}{\Gamma(c-a_1)\Gamma(c-b)},\label{rn3}
\end{eqnarray}
\begin{eqnarray}
A_2=A_-[2M(1-4aj)]^{-(\beta+D_*-2)}\frac{\Gamma(c)\Gamma(a_1+b-c)}{\Gamma(a_1)\Gamma(b)}.\label{rn4}
\end{eqnarray}

Now let us turn to the far field region. Assuming that
$r\rightarrow \infty$ and keeping only the dominant terms, we can
expend the wave equation (\ref{radial}) for the massless scalar
field as a power series in $1/r$
\begin{eqnarray}
\frac{d^2R_{FF}(r)}{dr^2}+\frac{3}{r}\frac{dR_{FF}(r)}{d
r}+\bigg[\tilde{\omega}^2-\frac{4l(l+1)}{r^2}\bigg]R_{FF}(r)=0,
\end{eqnarray}
with
\begin{eqnarray}
\tilde{\omega}^2=\omega^2-8\lambda j\omega.\label{ws0}
\end{eqnarray}
This is a Bessel equation. Thus, the general solution of radial
master equation (\ref{radial}) in the far field region can be
expressed as
\begin{eqnarray}
R_{FF}(r)=\frac{1}{r}\bigg[B_1J_{\nu}(\tilde{\omega}r)+B_2Y_{\nu}
(\tilde{\omega}r)\bigg],\label{rf}
\end{eqnarray}
where $J_{\nu}(\tilde{\omega}r)$ and $Y_{\nu}(\tilde{\omega}r)$
are the first and second kind Bessel functions, $\nu=2l+1$. $B_1$
and $B_2$ are integration constants. We now extend the far-field
asymptotic solution (\ref{rf}) towards small radial coordinate.
Taking the limit $r\rightarrow 0$, we obtain
\begin{eqnarray}
R_{FF}(r)\simeq\frac{B_1(\frac{\tilde{\omega}r}{2})^{\nu}}{r\;\Gamma(\nu+1)}
-\frac{B_2\Gamma(\nu)}{\pi
r\;(\frac{\tilde{\omega}r}{2})^{\nu}}.\label{rfn2}
\end{eqnarray}
Comparing equations (\ref{rn2}) and (\ref{rfn2}) in the low energy
and low angular momentum limit, we can obtain two relations
between $A_1$ and $B_1,\;B_2$ in the limit $\omega r_+\ll1$.
Making use of equations (\ref{rn3}) and (\ref{rn4}) and removing
$A_-$, we find the constraint for $B_1,\; B_2$
\begin{eqnarray}
B\equiv\frac{B_1}{B_2}=-\frac{1}{\pi}\bigg[\frac{2}{M(1-4aj)\tilde{\omega}^2}\bigg]^{2l+1}
(2l+1)\frac{\Gamma^2(2l+1)
\Gamma(c-a_1-b)\Gamma(a_1)\Gamma(b)}{\Gamma(a_1+b-c)\Gamma(c-a_1)\Gamma(c-b)}.
\label{BB}
\end{eqnarray}
In the asymptotic region $r\rightarrow \infty$, the solution in the
far field can be expressed as
\begin{eqnarray}
R_{FF}(r)\simeq
\frac{B_1+iB_2}{\sqrt{2\pi\tilde{\omega}}r^{\frac{3}{2}}}e^{-i\tilde{\omega}r}+
\frac{B_1-iB_2}{\sqrt{2\pi\tilde{\omega}}r^{\frac{3}{2}}}e^{i\tilde{\omega}r}=
A^{(\infty)}_{in}\frac{e^{-i\tilde{\omega}r}}{r^{\frac{3}{2}}}
+A^{(\infty)}_{out}\frac{e^{i\tilde{\omega}r}}{r^{\frac{3}{2}}}.\label{rf6}
\end{eqnarray}
We need to point out that only when the condition
$\tilde{\omega}\geq0$ (i.e, $\omega\geq 8\lambda j$ ) is
satisfied, the solution (\ref{rf6}) denotes an incoming and an
outgoing spherical waves at large distance from the black hole.

The absorption probability can be calculated by
\begin{eqnarray}
|\mathcal{A}_{lm\lambda}|^2=1-\bigg|\frac{A^{(\infty)}_{out}}{A^{(\infty)}_{in}}\bigg|^2
=1-\bigg|\frac{B-i}{B+i}\bigg|^2=\frac{2i(B^*-B)}{BB^*+i
(B^*-B)+1}.\label{GFA}
\end{eqnarray}
Combining the above result and the expression $B$ given in
equation (\ref{BB}), we can analyze the properties of absorption
probability for the massless scalar field in a rotating G\"{o}del
black hole background in the low-energy and low-angular momentum
limits.

\section{The absorption probability and Hawking radiation in the rotating G\"{o}del black hole}

We are now in a position to compute the absorption probability and
discuss Hawking radiation of a Kerr black hole embedded in the
G\"{o}del universe.

In Fig.(1), we examine the influence of the angular momentum $a$ of
black hole and G\"{o}del parameter $j$ on the absorption
probability. In the left figure, we plot the absorption probability
for the first partial waves ($l=0, \;m=0, \;\lambda=0$) by fixing
$j=0.2$. One can easily see that the absorption probability
decreases with the increase of the parameter $a$, which is similar
to that in the general rotating black hole spacetime shown in
\cite{Kanti,Kan1,Kan2,Kan3,Kan4,Kan5,Kan6,Haw3,Haw4,Haw5}. For the
fixed $j$, we observed that for smaller $a$, the potential peak
becomes lower, which allows more radiation to leak to the infinity.
In the right figure, we fix $a$ (with constant angular momentum )
and exhibit the dependence of the absorption probability on the
G\"{o}del parameter $j$. It shows that with the increase of $j$, the
absorption probability decreases. In the low-energy and low-angular
momentum limit the absorption probability
$|\mathcal{A}_{lm\lambda}|^2\sim
\omega^3\;r^3_+[1+2Ma^2/r^4_+]^{\frac{3}{2}}$. The radius of the
black hole event horizon decreases with the increase of the
parameters $j$ and $a$. Thus for the first partial waves the
influence of the G\"{o}del parameter $j$ on the absorption
probability is similar to that of the angular momentum $a$ of the
black hole.

\begin{figure}[ht]
\begin{center}
\includegraphics[width=8.0cm]{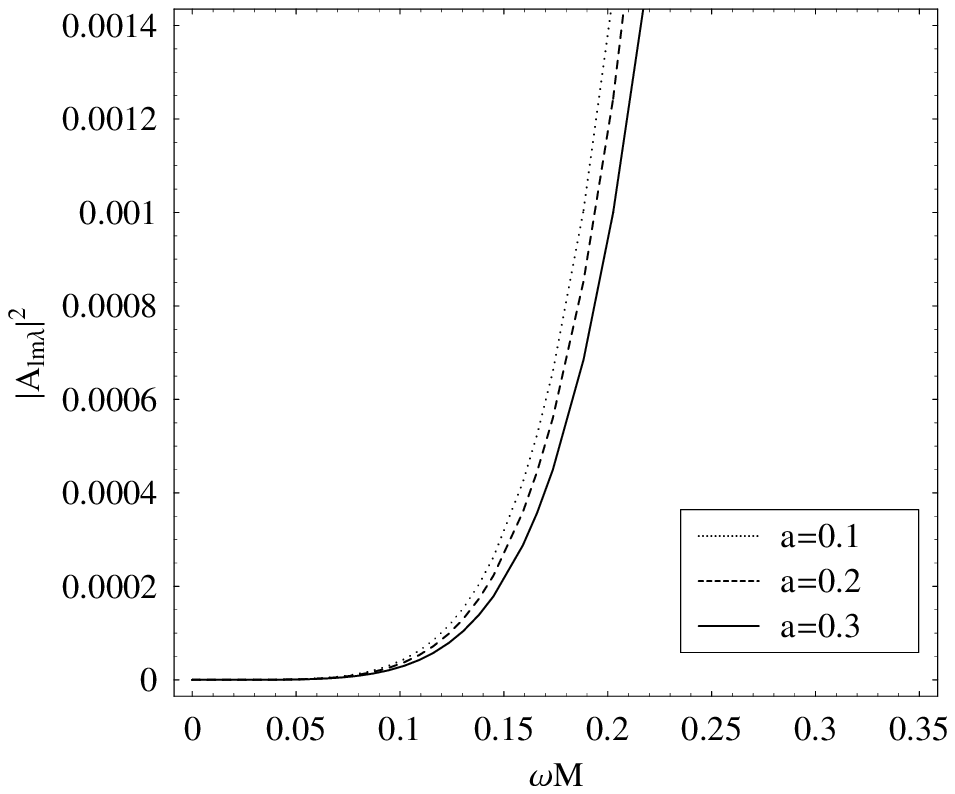}\;\;\;\;\includegraphics[width=8.0cm]{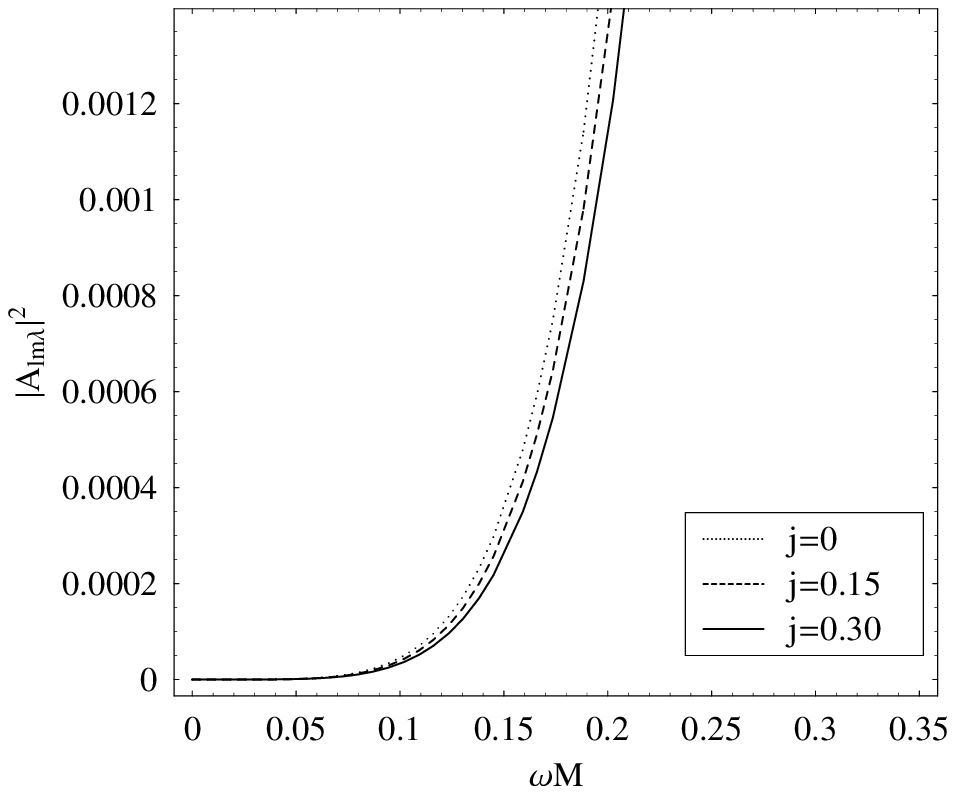}\;\;\;\;\;
\caption{Variety of the absorption probability
$|\mathcal{A}_{lm\lambda}|^2$ of
 scalar particles propagating in the rotating
G\"{o}del Black Holes in Minimal Five-Dimensional Gauged
Supergravity with $a$ (the left $j=0.2$) and with $j$ (the right
$a=0.15$), for fixed $l=0$, $\lambda=0$. }
 \end{center}
 \label{fig1}
\end{figure}

\begin{figure}[ht]
\begin{center}
\includegraphics[width=8.cm]{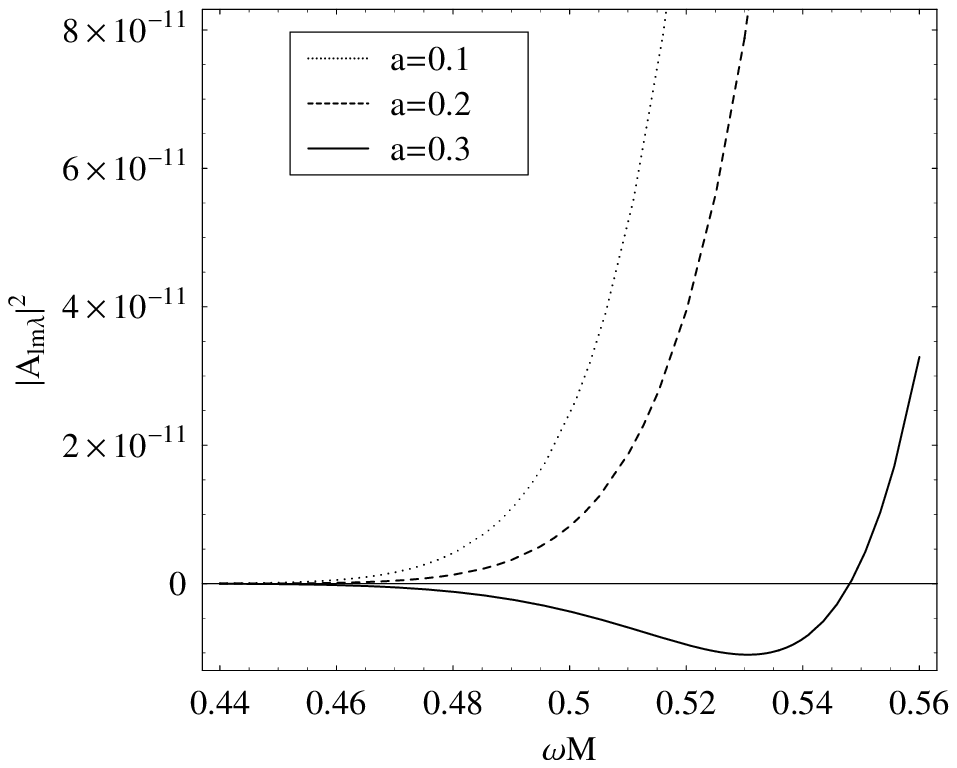}\;\;\;\;\;\includegraphics[width=8.2cm]{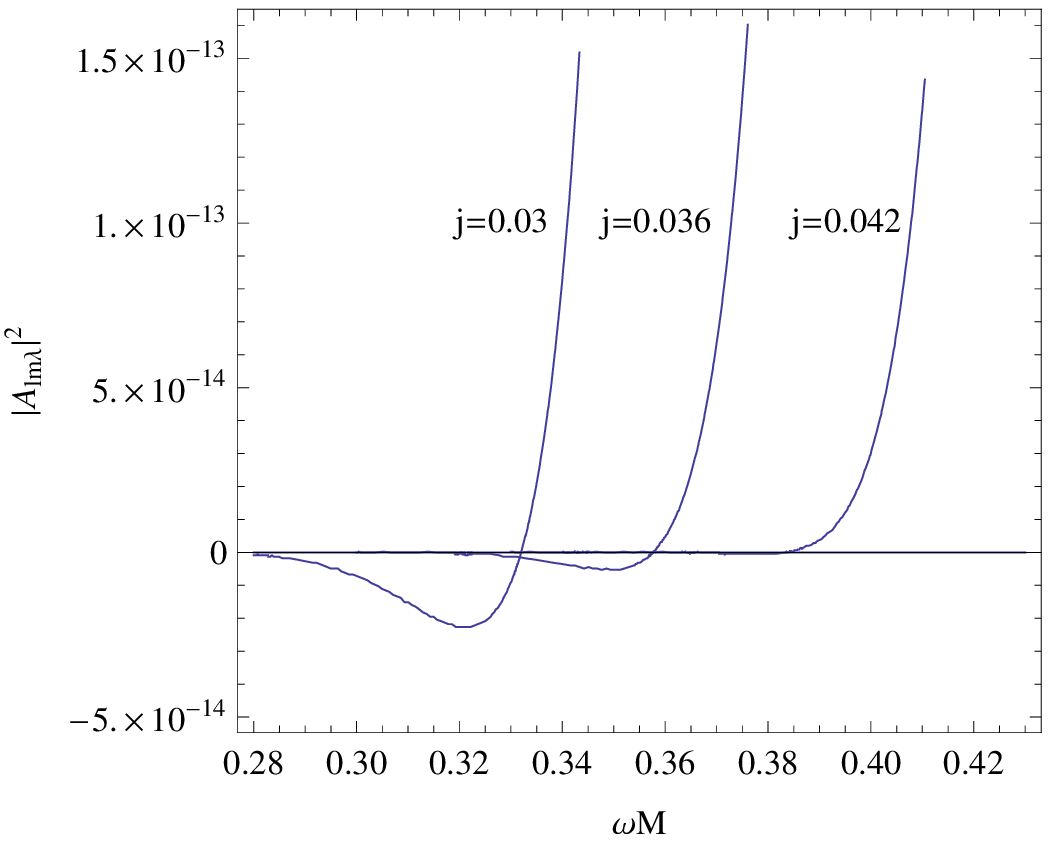}
\caption{Absorption probability $|\mathcal{A}_{lm\lambda}|^2$ of
scalar particles propagating in the rotating in the rotating
G\"{o}del Black Holes in Minimal Five-Dimensional Gauged
Supergravity, for $l=1$, $\lambda=1$. The left is for fixed $j=0.05$
and different $a$,  and the right is for fixed $a=0.2$ and different
$j$.}
\end{center}
\label{fig2}
\end{figure}

In Fig.(2), we find that for positive $\lambda$, in some range of
the frequency $\omega$, the absorption probability can be
negative, which presents us the super-radiance. This is the
property brought by the rotation as also disclosed in
\cite{Kanti,Kan1,Kan2,Kan3,Kan4,Kan5,Kan6,Haw3,Haw4,Haw5}. Besides
the effect of $a$, we observed that the G\"{o}del parameter $j$
can also contribute to the super-radiation. For the fixed $a$, We
find that with the increase of $j$, both the magnitude of the
super-radiance and the range of $\omega$ for the super-radiance to
happen decreases. It is interesting to see that the effect of the
G\"{o}del parameter $j$ on the super-radiance is different from
that caused by the angular momentum $a$ of the black hole. This
property of the G\"{o}del parameter $j$ has not been observed
elsewhere.

As in \cite{Haw3}, in the low energy limit $BB^*\gg i(B^*-B)\gg 1$,
we can simplify our (\ref{GFA}) to the form
\begin{eqnarray}
|\mathcal{A}_{lm\lambda}|^2&=&2i(\frac{1}{B}-\frac{1}{B^*}) \no\\
&=&4\pi
\bigg[\frac{M(1-4aj)\tilde{\omega}^2}{2}\bigg]^{2l+1}\frac{K*}{A(r_+)}
\frac{\Gamma^2(2\beta+D_*-2)\Gamma^2(1-\beta)(2-D_*-2\beta)\sin^2\pi(2\beta+D_*)}{(2l+1)\Gamma^2
(2l+1)\Gamma^2(\beta+D_*-1)\sin^2\pi(\beta+D_*)}.
\end{eqnarray}
From (\ref{bet}) we learnt that the quantity $2-D_*-2\beta$ is
always positive. Using (\ref{ax}) we have
$A(r_+)=1-\frac{2Ma^2}{r^4_+}$, which is positive since
$r^2_+-\sqrt{2M}a=M(1-4aj)-\sqrt{2M}a+\sqrt{M^2(1-4aj)^2-2Ma^2}>0$.
$\tilde{\omega}\geq0$ is required to describe the outgoing and
incoming spherical waves in the large distance in (\ref{rf6}). The
possibility to make $|\mathcal{A}_{lm\lambda}|^2<0$ is $K_*<0$. From
(\ref{kx}) and (\ref{ws0}), $K_*<0$ and $\tilde{\omega}\geq0$ lead
to
\begin{eqnarray}
0\leq \omega\leq \omega_c=\frac{4\lambda (jr^4_++Ma)}{r^4_++2Ma^2},
\label{wc}
\end{eqnarray}
and
\begin{eqnarray}
\omega\geq\omega_0=8\lambda j\;,\label{w0}
\end{eqnarray}
respectively. The condition for the occurrence of the
super-radiance in this black hole background is
$\omega_0\leq\omega_c$. From (\ref{wc}) and (\ref{w0}), we obtain
the ratio
\begin{eqnarray}
\frac{\omega_0}{\omega_c}=\frac{2j(r^4_++2Ma^2)}{jr^4_++Ma}.
\end{eqnarray}
\begin{figure}[ht]
\begin{center}
\includegraphics[width=8.0cm]{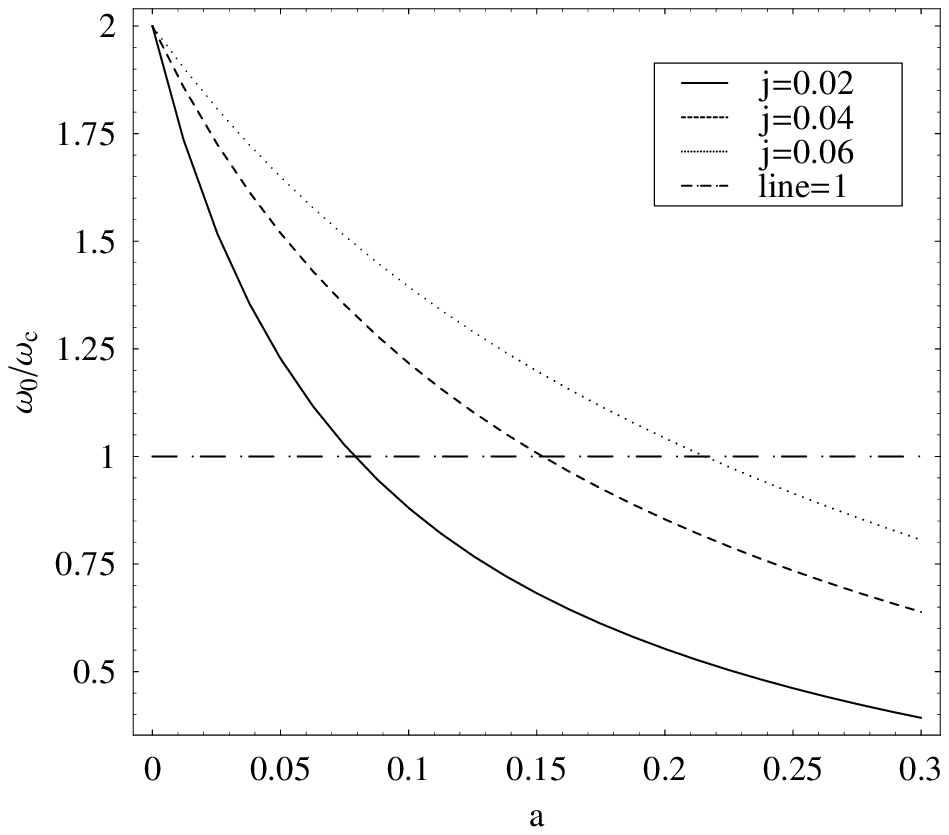}\;\;\;\;\;\;\includegraphics[width=8.0cm]{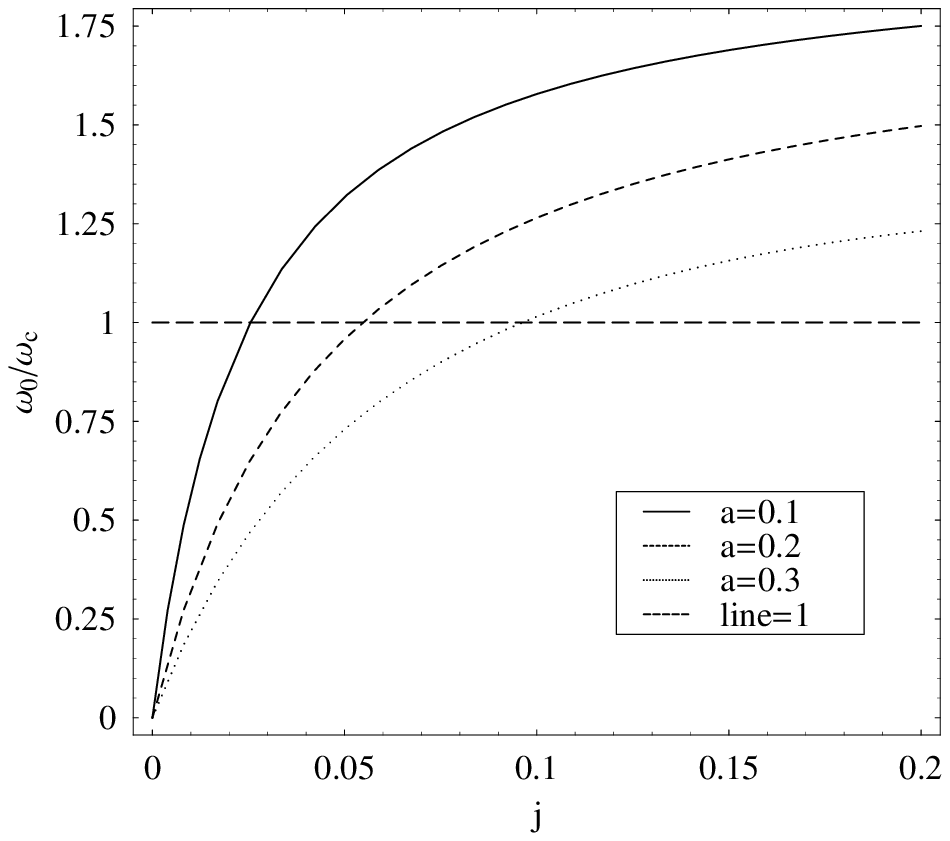}
\caption{Ratio of the value $\omega_0/\omega_c$ with the change of
$a$ and $j$.}
\end{center}
\label{fig3}
\end{figure}
When $a=0$, from the above formula we have $\omega_0=2\omega_c$.
This means that although the Schwarzschild G\"{o}del black hole
contains the rotational parameter $j$ of the G\"{o}del universe it
is impossible to occur super-radiance.

Fig.(3) shows the changes of the ratio $\omega_0/\omega_c$ with
$a$ and $j$, which tells us that for the fixed $j$, there exists a
lower bound of $a$ for the super-radiance to occur. But for the
fixed $a$, there is the upper bound of $j$ for the super-radiance
to happen. This is also shown in Fig.(2).

\begin{figure}[ht]
\begin{center}
\includegraphics[width=8.0cm]{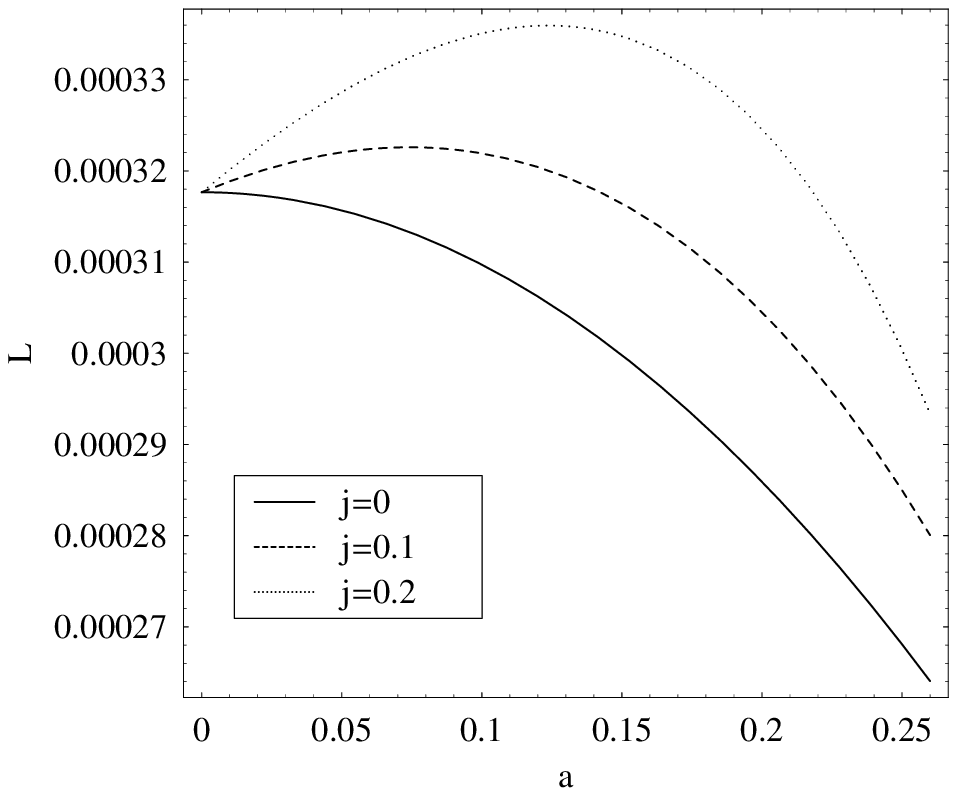}\;\;\;\;\; \includegraphics[width=8.0cm]{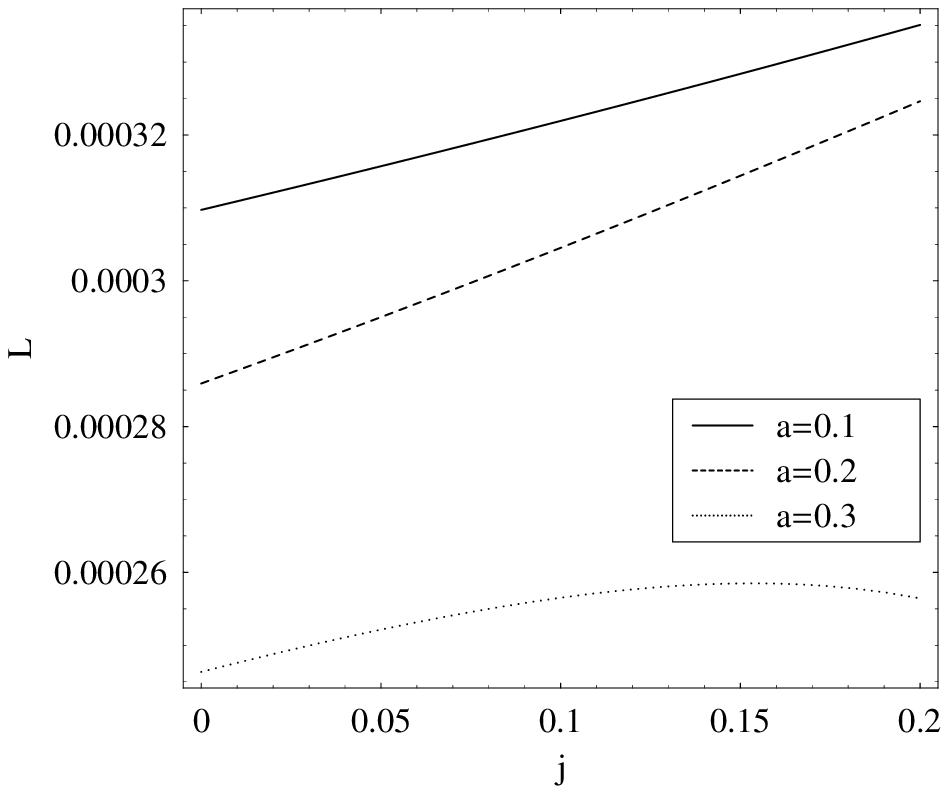}
\caption{The luminosity of Hawking radiation $L$ of scalar particles
propagating in the rotating G\"{o}del Black Holes in Minimal
Five-Dimensional Gauged Supergravity, for $l=0$ and $\lambda=0$ and
different $j$ and $a$.}
\end{center}
\label{fig4}
\end{figure}

Now let us turn to study the luminosity of the Hawking radiation for
the mode $l=0$, $\lambda=0$ which plays a dominant role in the
greybody factor. Performing an analysis similar to that in
\cite{Haw3,Haw4,Haw5}, we can rewrite the greybody factor
(\ref{GFA}) as
\begin{eqnarray}
|\mathcal{A}_{lm\lambda}|^2&\simeq&
\frac{\pi\omega^3\;r^3_+}{2}\bigg[1+\frac{2Ma^2}{r^4_+}\bigg]^{\frac{3}{2}}.\label{GFA1}
\end{eqnarray}
For the slowly rotating black hole, when $j$ increases, the greybody
factor decreases. In the limit $a\rightarrow 0$, the form of the
formula (\ref{GFA1}) reduces to that in the usual five dimensional
Schwarzschild black hole spacetime \cite{TJR}. Combining it with
equation (\ref{T1}), the luminosity of the Hawking radiation is
given by
\begin{eqnarray}
L&=&\int^{\infty}_0\frac{d\omega}{2\pi}
|\mathcal{A}_{lm\lambda}|^2\frac{\omega}{e^{\;(\omega-\Omega_H\lambda)/T_{H}}-1}\nonumber\\
&\simeq&\frac{3\zeta(5)}{16\pi^5}\frac{r^2_+}{r^4_++2Ma^2}\bigg(1-\frac{r^2_-}{r^2_+}\bigg)^5
=GT^5_H, \label{LHK}
\end{eqnarray}
where
$G=6\zeta(5)r^3_+\bigg[1+\frac{2Ma^2}{r^4_+}\bigg]^{\frac{3}{2}}$.
\begin{table}[h]
\begin{center}
\begin{tabular}[b]{cccccc}
 \hline \hline
 \;\;\;\; $j$ \;\;\;\; & \;\;\;\; 0 \;\;\;\; & \;\;\;\; 0.05\;\;\;\;
 & \;\;\;\; 0.1 \;\;\;\;& \;\;\;\; 0.15 \;\;\;\; & \;\;\;\; 0.2\;\;\;\; \\ \hline
\\
$a_p$& \;\;\;\;\;0\;\;\;\;\;  & \;\;\;\; 0.0392\;\;\;\;\;
 & \;\;\;\;\;0.0742\;\;\;\;& \;\;\;\;\;0.1025\;\;\;\;\; & \;\;\;\; 0.1238\;\;\;\;\;
\\ \hline\hline
\end{tabular}
\caption{The change of $a_p$ with different $j$. Here $M=1$. }
\end{center}
\end{table}
In figure (4), we show the dependence of the luminosity of Hawking
radiation on the G\"{o}del parameter $j$ for different angular
momentum parameters. In the limit $j\rightarrow 0$, the luminosity
of Hawking radiation $L$ decreases with the increase of $a$, which
is consistent with that in usual five dimensional Kerr black hole
spacetime. However, when $j\neq 0$, from (\ref{LHK}) we find that
there is a peak of the luminosity of the Hawking radiation when
$dL/da=0$. For fixed $M=1$ but different $j$, peaks appear at
different $a_p$ which are listed in table I. With the increase of
$j$, the $a_p$ increases. When $a<a_p$, we observe in Fig.(4) that
with the increase of $a$, $L$ increases.  For $a>a_p$, we see in
Fig.(4) that the luminosity of Hawking radiation decreases with the
further increase of $a$. When $a\rightarrow 0$, the formula
(\ref{LHK}) reduces to that in the five dimensional Schwarzschild
G\"{o}del black hole spacetime. In the small $a$ case, we find that
the luminosity of the Hawking radiation $L$ increases with $j$,
which can be explained by Hawking temperature of the black hole. In
the small $a$ regime the behavior of the Hawking temperature $T_H$
can be expressed as
\begin{eqnarray}
T_H\sim\frac{1}{\sqrt{2M}}+\frac{\sqrt{2}j}{\sqrt{M}}\;a
+\frac{12j^2M-1}{\sqrt{2M}}\;a^2+O(a^3).
\end{eqnarray}
Thus, as $j$ increases the Hawking temperature increases. This leads
to that the luminosity of Hawking radiation $L$ increases with $j$
in the small $a$ case. Our result also implies that the G\"{o}del
parameter can enhance the Hawking radiation. The influence of the
G\"{o}del parameter on the luminosity of Hawking radiation is in
agreement with that observed in the quasinormal modes results in
\cite{qu1}. The larger G\"{o}del parameter $j$ enhances the power
emission of the black hole so that it is more difficult for the
perturbation outside the black hole to die out.

\section{summary}
In this paper, we have studied the greybody factor and Hawking
radiation for a massless scalar field in the background of a five
dimensional rotating G\"{o}del black hole in the low energy and
low angular momentum approximation. We have found that the
absorption probability and Hawking radiation contain the imprint
of the G\"{o}del parameter. With the inclusion of the G\"{o}del
parameter $j$, we have observed richer physics which has not been
shown in the usual rotating black hole background. The effects on
the super-radiance and the luminosity of the Hawking radiation due
to the G\"{o}del parameter $j$ are different from that of the
angular momentum of the black hole. The observation that the
G\"{o}del parameter can enhance the Hawking radiation is
interesting and this might open a window to detect whether our
universe is rotating or not. It would be of interest to generalize
our study to other fields emission, such as the gravitational
field etc. Work in this direction will be reported in the future.

\begin{acknowledgments}

This work was partially supported by NNSF of China, Shanghai
Education Commission and Shanghai Science and Technology Commission.
S. B. Chen's work was partially supported by the Scientific Research
Fund of Hunan Provincial Education Department Grant No. 07B043 and
the construct program of key disciplines in Hunan Province. J. L.
Jing's work was partially supported by the National Natural Science
Foundation of China under Grant No. 10675045; the FANEDD under Grant
No. 200317; and the Hunan Provincial Natural Science Foundation of
China under Grant No. 07A0128.
\end{acknowledgments}

\end{document}